\documentclass[graybox]{svmult}

\pdfoutput=1

\usepackage{mathptmx}       
\usepackage{helvet}         
\usepackage{courier}        
\usepackage{type1cm}        
\usepackage{makeidx}         
\usepackage{graphicx}        
\usepackage{multicol}        
\usepackage[bottom]{footmisc}

\usepackage{amssymb}

\usepackage{epstopdf}

\usepackage{caption}
\usepackage{subcaption}
\makeindex             

\begin{document}

\title*{Evacuation Dynamics Of Asymmetrically Coupled Pedestrian Pairs}
\author{Frank M\"uller and Andreas Schadschneider}
\institute{Frank M\"uller \and Andreas Schadschneider \at
Institut f\"ur Theoretische Physik, Universit\"at zu K\"oln, 50937 K\"oln,
Germany, \email{fm@thp.uni-koeln.de, as@thp.uni-koeln.de} 
}
%
%
\maketitle

\abstract{We propose and analyze extended floor field cellular
  automaton models for evacuation dynamics of inhomogeneous pedestrian
  pairs which are coupled by asymmetric group interactions. Such pairs
  consist of a leader, who mainly determines the couple's motion and a
  follower, who has a defined tendency to follow the leader. Examples
  for such pairs are mother and child or two siblings of different
  age. We examine the system properties and compare them to the case
  of a homogeneous crowd. We find a strong impact on evacuation times
  for the regime of strong pair coupling due to the occurrence of a
  clogging phenomenon. In addition we obtain a non-trivial dependence
  of evacuation times on the followers' coupling to the static floor
  field, which carries the information of the shortest way to the exit
  location. In particular we find that systems with fully passive
  followers, who are solely coupled to their leaders, show lower
  evacuation times than homogeneous systems where all pedestrians have
  an equal tendency to move towards the exit. We compare the results
  of computer simulations with recently performed experiments.}


\section{Introduction}
\label{Intro}
Human crowds and pedestrian traffic are usually composed of both
social groups and individuals. Recent empirical studies brought to
attention that in this context social groups are rather the normality
than the exception \cite{Proceedings_TGF15}.  Moussa\"{\i}d et al.
\cite{Moussaid} observed in a field study that up to 70\% of
pedestrians walk in social groups and Xi et al. \cite{JianXi} found
that most pedestrians walk in two-person-groups whereas individual
pedestrian traffic is only second frequent.  The high relevance of
two-person-groups shows the importance of a deeper understanding of
the impact such groups impose on evacuation processes.

We recently introduced models for evacuation processes including
social groups which are inspired by methods of non-equilibrium physics
\cite{Mueller2014}. From the perspective of physics the pivotal
characteristic of these models is that social groups are
\mbox{\textit{cohesive}}, i.e.\ group members tend to maintain a
spatial coherence. The current work focuses on asymmetrically coupled
two-person-groups and aims to provide models for such pairs in
evacuation processes. Being the smallest social group
two-person-groups still can be very diverse. The strength and symmetry
of interaction as well as the group members' level of orientation can
differ significantly. Thus it has to be verified which set of models
is applicable for simulations. The model types studied here can e.g.
describe pairs like mother and child or siblings of different age
where one part will dominantly determine the motion and the other part
will have a defined tendency to follow. The dominating part will be
called \textit{leader} and the following part will be called
\textit{follower}. The proposed models are used for computer
simulations and characteristic effects will be discussed.

The basic underlying model used for the computer simulations is the
floor field cellular automaton model (FFCA). It is a stochastic model
defined on a \mbox{2-dimensional} grid with time evolution in discrete
steps. A cell can be either empty or occupied by a particle representing
a pedestrian. Particles can move by transition to
a neighboring cell.  The transition will take place with a transition
probability arising from different floor fields which encode the
tendency to move towards the room's exit and the interaction between
pedestrians - here the group cohesion. Further details and general
properties of the FFCA can be found in \cite{Mueller2014}.

The evacuation simulations are performed on a standard grid of $ 63
\times 63 $ cells with a moderate pedestrian density of $ \rho = 0.02
$. Observables are averaged over at least 500 runs.


\section{The DGFF and MTFF as mediators of group interaction}
\label{Generators}

The DGFF and MTFF are the central components which create the group
cohesion in the models studied here. We introduced both concepts in
\cite{Mueller2014} and will recap the most important properties in the
following.


\subsection{Properties of the DGFF}
\label{DGFF1}
In the first proposed model the DGFF mediates the interaction between
a leader and the follower. A FFCA with DGFF provides a model for
crowds with two-person-groups which have a bond with a likelihood to
permanently break up in higher densities.

The DGFF extends the dynamic floor field (DFF) introduced in
\cite{Kirchner} in several respects. It shares the basic idea that
pedestrians increase a field value in the cell they leave when moving
to a neighboring cell while decay and diffusion can modify it over
time. This accounts for the important property that the DGFF is not
related to position, but to the movement the leader has performed in
preceding time steps. Each couple interacts via its individual DGFF.
Only when the leader moves the field is built up and only the follower
who is associated with the respective leader reacts on this leader's
DGFF. If we do not take the SFF into account followers are most likely
to transit to cells with a high associated DGFF value. This causes
group members to tend moving on the same trajectory. The full
transition probability for the follower including the SFF and the DGFF is
\begin{equation} 
p_{ij}^{F(s)}=N \exp\left(k_{S}^{F}S_{ij}\right)
\exp\left(k_D^{F} D_{ij}^{F(s)}\right) (1-\eta_{ij}) \xi_{ij}\,.
\label{eq_transprobFollower}
\end{equation}
Here $p_{ij}^{F(s)}$ is the transition probability for the follower
of pair $s$, $k_{S}^{F}$ is the coupling constant to the static
floor field for followers and $k_D^{F}$ is the coupling constant to
the DGFF for followers, determining the coupling strength to this
field. $N$ is a normalization term. The product $ (1-\eta_{ij})\xi_{ij}$ 
guarantees the exclusion principle and avoids transition into wall
cells, see e.g.\ \cite{Kirchner}.

The transition probability for the leader only considers the SFF,
which encodes the shortest way to the exit:
\begin{equation} 
p_{ij}^{L(s)}=N \exp(k_{S}^{L}S_{ij})(1-\eta_{ij})\xi_{ij}\,.
\label{eq_transprobLeader}
\end{equation} 

The DGFF can be understood as a field composed of field quanta
(bosons\footnote{Despite the denomination, "bosons" should not be
  considered as quantum mechanical particles.}) with defined internal
degrees of freedom. E.g. each field quantum carries the information by
which particle it was produced.  This allows particles to interact
only with bosons of a special type.  In particular it enables a
follower particle to ignore all bosons, but these of its leader. The
leader particles do not react with any boson type. This way an
asymmetric group interaction can be established while self-interaction
of particles is completely avoided.

It is an important characteristic of the DGFF concept that a moving
particle increases the DGFF by $ m\gg 1$ instead of only $ m=1$ as it
is the case in \cite{Kirchner}. Small values of $m$ would not lead to
sufficiently structured boson traces the followers could continuously
follow since only one particle will contribute to the DGFF whereas in
case of the DFF all particles contribute to the field. In addition a
diffusion parameter $\alpha > 0$ is important for continuous group
cohesion as it broadens the boson trace. Both factors highly increase
the probability that followers do not lose the tracks of their
leaders.  \mbox{Figure}~\ref{fig_numbosons} illustrates the strong
dependency of the average distance between leaders and followers $d$
on the diffusion parameter $ \alpha $ and the boson multiplicity $m$. 
$d$ is a measure for group cohesion. While $m=1$ does not
create any noticeable pair bond, $m=400$ causes strong group cohesion.


\subsection{Construction of the Moving Target Floor Field}
\label{sec_MTFF1}
As explained in Section~\ref{DGFF1} the DGFF is solely increased in a
cell when the leader leaves the cell by moving to a neighboring cell.
Thus the DGFF is depending on the leader's movement. The question
arises how evacuation dynamics changes when the underlying floor field
is solely depending on the leader's position. This is the case with the
second model we examine here - the moving target floor field (MTFF).

In the FFCA model with MTFF group cohesion is achieved by an
asymmetric interaction related to the relative position of the leader
with respect to the follower.  Every leader of pair $ s $ induces a
group-specific floor field $ M^{(s)} $ in the von Neumann neighborhood
of his associated follower:
\begin{equation} M^{(s)}_{ij}(T)=\max_{(\tilde i, \tilde j)} 
\biggl\{\sqrt{ (i_L(T)- \tilde i )^2+(j_L(T)- \tilde j )^2 } \biggr\} 
- \sqrt{ (i_L(T)-i)^2+(j_L(T)-j)^2 }
\end{equation}
$ (i_L(T), j_L(T)) $ denotes the position of the leader at time step $
T $.  The first term on the r.h.s. is a normalization term where $
(\tilde i, \tilde j) $ runs over the cells in the von Neumann
neighborhood of the follower.

The MTFF contributes to the transition probabilities in an analogous
manner as the DGFF in Section~\ref{DGFF1}. The total transition
probability of the follower $F$ of pair $s$ is
\begin{equation}
p_{ij}^{F(s)}=N \exp((k_{S}^{F}S_{ij}+k_M M^{(s)}_{ij})(1-\eta_{ij})
\xi_{ij} \,.
\end{equation}


\section{Impact of pair cohesion on evacuation dynamics}
\label{sec_SimulationResults}

In this section we will investigate the question how the fragmentation
of a pedestrian crowd into asymmetrically coupled pedestrian pairs
impacts evacuation dynamics and how the resulting evacuation process
compares to the scenario with a homogeneous crowd without pair bonds.
We will analyze both DGFF and MTFF systems and compare the results.



\subsection{Comparison of DGFF, MTFF and homogeneous model}
\label{sec_Comparison}
For the purpose of comparison with a homogeneous crowd the model is
configured such that leaders and followers are equipped with the same
level of orientation, which is realized by an equal coupling constant
with respect to the SFF. When \mbox{$k_{D}^{F} = 0$} the
configuration $ k_{S}^{L} = k_{S}^{F} $ coincides with a homogeneous
crowd with no interaction between the pedestrians.

\begin{figure}[htbp]
  \centering
  \begin{minipage}[c]{5.5 cm}
  \center
\resizebox{1.0\textwidth}{!}{\includegraphics{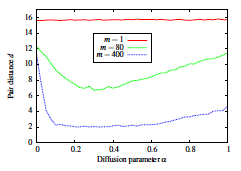}}    
    \vspace{-3mm}
    \subcaption{}
    \label{fig_numbosons}
  \end{minipage}
   \hfill
  \begin{minipage}[c]{5.5 cm}
    \center
\resizebox{1.0\textwidth}{!}{\includegraphics{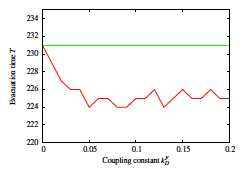}}  
  \vspace{-3mm}
    \subcaption{}
    \label{fig_Pairs_DGFF_kd2}
\end{minipage}
\caption{Different views on the DGFF model. 
  \textbf{(a)} Dependence of the average pair distance $ d $ on the
  diffusion parameter $ \alpha $ for different values of the boson
  multiplicity $m$. 
  \textbf{(b)} Dependence of $ T $ on $ k_{D}^{F}$. The dashed line 
  refers to $ T = T_{hom} $, which is the average evacuation time of 
  a homogeneous crowd without pair bonds.}
\label{DGFF_Evactime}
\end{figure}

First we turn to the model with DGFF. For growing $ k_{D}^{F} $ the
homogeneous crowd is fragmented into asymmetrically coupled pairs
which increasingly maintain proximity and tend to move on the same
trajectory. How will this impact evacuation dynamics?
Figure~\ref{fig_Pairs_DGFF_kd2} shows the dependence of $ T $ on $
k_{D}^{F} $ for the DGFF model: The coupling in pairs slightly
improves the evacuation process. The effect is small, but it is
clearly visible that $T$ drops for growing $ k_{D}^{F} $. The drop
takes place in a comparably small interval since the boson
multiplicity is high at $m=400$. It is interesting to note that the
effect does not coincide with a lower average number of conflicts per
time step. An analysis of this number shows that conflicts are even
increased for $ k_{D}^{F} > 0$, but still $T$ is lowered. The higher
number of conflicts is due to the group cohesion. The continuous
proximity of group members increases the likelihood that these choose
the same cell for a transition which leads to an overall increase of
conflicts.

The improvement of evacuation time $ T $ in this model becomes
comprehensible when recalling the nature of the DGFF. It encodes
spatio-temporal information about the path the leader has successfully
moved on - not only spatial information about the leader. Without
movement no DGFF builds up. Therefore the information contributes to
choose {\em successful} paths through the crowd.

\begin{figure}[htbp]
  \centering
  \begin{minipage}[c]{5.5 cm}
  \center
\resizebox{1.0\textwidth}{!}{\includegraphics{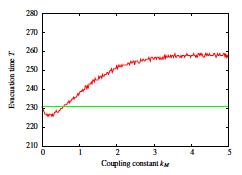}}
    \vspace{-3mm}
    \subcaption{}
    \label{MTFF_equal_ks}
  \end{minipage}
   \hfill
  \begin{minipage}[c]{5.5 cm}
    \center
\resizebox{1.0\textwidth}{!}{\includegraphics{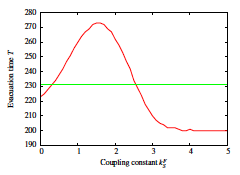}}  
    \vspace{-3mm}
    \subcaption{}
    \label{img_k_s}
  \end{minipage}
\caption{Different views on the average evacuation time $T$ 
  for systems with MTFF. The dashed line visualizes $ T = T_{hom} $,
  which is the average evacuation time of a homogeneous crowd without
  pair bonds and $k_S=0.8$. \textbf{(a)} Dependence of evacuation time
  $ T $ on the coupling strength $ k_M $ in a configuration with $
  k_{S}^{L}= k_{S}^{F}=0.8 $. \textbf{(b)} Dependence of evacuation
  time $ T $ on $k_{S}^{F}$ which controls the follower's coupling to
  the SFF. Pair coupling strength is constant at $ k_M = 2 $ and $
  k_{S}^{L}=0.8 $.}
\label{DGFF_Evactime2}
\end{figure}

The coupling mechanism of the MTFF highly differs from that of the
DGFF since it is not related to the movement of the leader, but to his
position. In fact simulations show that this difference translates to
measurable differences in the respective evacuation processes.
Figure~\ref{MTFF_equal_ks} is the analogon of
Figure~\ref{fig_Pairs_DGFF_kd2} for the MTFF model.  It shows a
non-trivial dependence of the evacuation time $ T $ on the coupling
parameter $ k_M $.  The dashed line in Figure~\ref{MTFF_equal_ks}
refers to the average evacuation time $ T $ of a homogeneous system
without pair bonds.  In both figures coupling to the SFF is at $
k_{S}^{L}=k_{S}^{F}=0.8 $. In the domain of $ k_M \lesssim 0.6 $ the
coupling in pairs results in a lower $ T $ while for $ k_M \gtrsim 0.6
$ $ T $ is increased. $ T $ remains nearly constant for $ k_M \gtrsim
4 $ ($ T \approx 258 $) until clogging processes increase $ T $ again.
Clogging is discussed in Section~\ref{sec_clogging}.

\subsection{Influence of the follower's coupling to the static floor field}
\label{sec_SFF}
In this section we shift the point of interest to the question how the
coupling of the follower to the static floor field (SFF) $ k_{S}^{F}$
influences the evacuation process. Apart from the coupling to the
leader via MTFF the follower is also coupled to the SFF, which encodes
the shortest way to the exit. $ k_{S}^{F} = 0 $ means that the
follower's motion is not oriented at the exit at all whereas $
k_{S}^{F} \rightarrow \infty $ leads to a deterministic movement on
the shortest path to the exit. $ k_{S}^{F}$ can be interpreted as the
follower's orientation towards the exit or more generally as the
ability and will to reach the exit himself.

Figure~\ref{img_k_s} shows the resulting average evacuation time for a
system with $ k_M = 2 $. The dashed line depicts the evacuation time $
T_{hom} $ of the homogeneous reference system without pair coupling
and equal $ k _S $. It is a counterintuitive result that fully passive
followers, who do not have any tendency to move to the exit themselves
lead to a more efficient evacuation with lower evacuation times than a
homogeneous crowd with an equally good orientation towards the exit.
It appears to be beneficial if followers are solely led by their
leaders. In contrary strong pair coupling together with the equally
good orientation towards the exit ($ k_{S}^{L}=k_{S}^{F}=0.8 $) slows
down the evacuation.  This result was also found in
Figure~\ref{MTFF_equal_ks}.

The domain of $ k_{S}^{F} \gtrsim 2.5 $ where $ T $ falls below $
T_{hom} $ again arises from a situation where followers have such good
ability to reach the exit themselves that despite their pair bond they
overtake their leaders and reach the exit first.  Here the overall
average evacuation time benefits from the fast evacuation of the
followers.


\subsection{Clogging and gridlocks}
\label{sec_clogging}
In \cite{Mueller2014} we had found clogging phenomena for the
asymmetric fixed-bond leader-follower model. In this model the pair
bond is fully fixed as for every cell transition the follower is
positioned on the cell the leader had occupied previously. The pair
distance is always $ d=1 $. The question remained if such clogging
phenomena can be found when the bond is dynamic and $d$ can take on
arbitrary values. Indeed this is the case for systems with MTFF.

\begin{figure}[b]
  \centering
  \begin{minipage}[c]{5.5 cm}
  \center
\resizebox{1.0\textwidth}{!}{\includegraphics{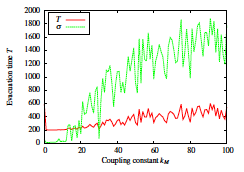}} 
    \vspace{-3mm}   
    \subcaption{}
    \label{img_clogging_T}
  \end{minipage}
   \hfill
  \begin{minipage}[c]{5.5 cm}
    \center
\resizebox{1.0\textwidth}{!}{\includegraphics{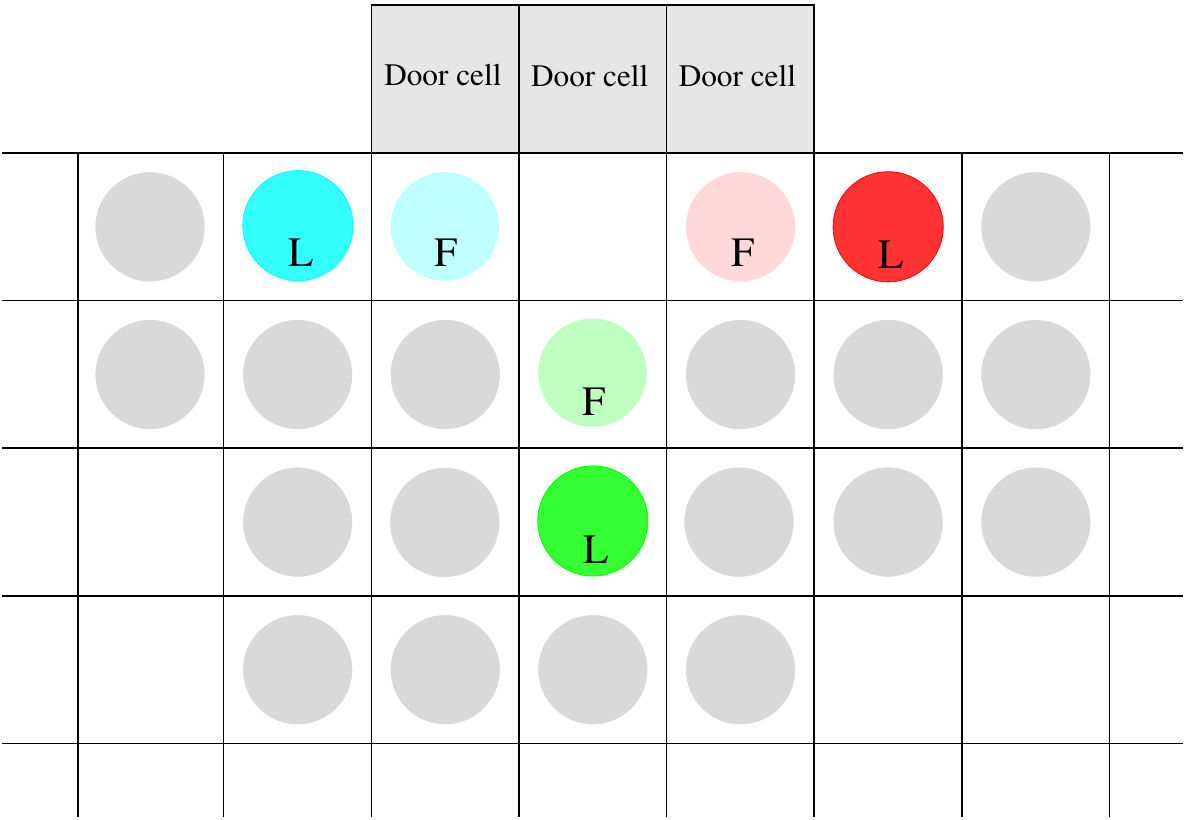}}  
    \vspace{-3mm}
    \subcaption{}
    \label{img_clogging}
  \end{minipage}
\caption{Clogging in systems with MTFF. 
  \textbf{(a)} Dependence of evacuation time $ T $ on the coupling
  strength $ k_M $ with $ k_{S}^{L}=0.8 $ and $ k_{S}^{F}=0.2 $. High
  values of $k_M$ lead to increased values of $T$ and high $
  \sigma(T)$ due to clogging. \textbf{(b)} Example for a typical
  clogging situation for a system with door width $d=3$.}
\label{DGFF_Evactime3}
\end{figure}

In Section~\ref{sec_Comparison} it was addressed that for high $k_M$
and the resulting high group cohesion clogging starts to increase the
average evacuation time $ T $.  Figure~\ref{img_clogging_T} shows such
increase. Here $ T $ fluctuates strongly and the standard deviation of
$ T $ is accordingly high. The phenomenon occurs due to followers who
maintain the nearest possible position to their leaders in front of
the exit and impede their leaders from reaching the exit. If a
configuration occurs where followers completely shield their leaders
from exiting no particle will be evacuated until the situation
dissolves. Figure~\ref{img_clogging} displays an example of such
situations.  For very high values of $k_M$ the shielding followers
have a nearly vanishing probability to ever leave the cell next to
their leader and it comes to a final stop of the evacuation
(gridlock). This is a situation which is not found in reality. However,
clogging due to pairs who do not let go each other in a high density
situation is well conceivable and a possible source of impediment in
evacuation processes through a narrow door. The high standard
deviation in the domain of clogging indicates that evacuation
scenarios become increasingly unpredictable once clogging becomes a
likely effect during the evacuation.

The described clogging phenomenon does not occur in systems with DGFF
as this field is solely built up by the motion of the respective
leader. While the MTFF only encodes spatial information about the
leader the DGFF encodes \mbox{spatio-temporal} information about the path the
leader has successfully moved on. When it comes to highly congested
states the DGFF is only rarely increased and the decay mechanism
brings the field strength down to low values or zero. Then movement is
mainly governed by the SFF and particles start moving towards the exit
again. Thus systems with DGFF cannot develop long-term clogging or
even gridlocks.


\section{Conclusion}
\label{sec_conclusion}

Quantitatively for moderate coupling both models show only small
deviations from the average evacuation time~$ T $ of a homogeneous
reference system without pair coupling. This result is in line with
the evacuation experiments we recently performed with students
\cite{VonKruechten}.  However, qualitatively the two models differ
significantly from each other as the DGFF model always leads to a
decreased $ T $ while the MTFF model shows two domains with decreased
$ T$ for low and increased $T$ for high coupling strength~$ k_M $. At
present the collected data from experiment does not provide sufficient
statistical significance to rank one model over the other. Further
experiments will contribute to investigate this question.

For the MTFF system our simulations have shown a non-trivial
dependency of~$ T $ on $ k_{S}^{F} $. The fragmentation of a crowd
into couples with fully passive followers results in a more efficient
evacuation process than a homogeneous crowd.

In the domain of strong coupling to the MTFF clogging leads to an
increase of~$ T $ and a high standard deviation, which makes the
average evacuation time $ T $ less meaningful and a single evacuation
process less predictable. This is an important factor when simulations
are to predict evacuation times, e.g. for evacuation assistants to
support decisions about optimum evacuation routes during
\mbox{emergencies}.



\begin{acknowledgement}
  We dedicate this contribution to the memory of our friend and
  colleague Matthias Craesmeyer. Financial support by the DFG under
  grant SCHA 636/9-1 is gratefully acknowledged.
\end{acknowledgement}


\bibliographystyle{unsrt}

\bibliography{MS_TGF15_bib}

\end{document}